\documentclass[aps,twocolumn,amsmath,amssymb,prb,showpacs,superscriptaddress]{revtex4-1}
\usepackage{amsfonts, hyperref, times}
\usepackage{graphicx}

\providecommand{\vect}[1]{{\boldsymbol{#1}}}

\begin{document}
  
\title{Effective description of domain wall strings
}

\author{Davi~R. Rodrigues}
\affiliation{Department of Physics \& Astronomy, Texas A\&M University, College Station, Texas 77843-4242, USA}
\affiliation{Institute of Physics, Johannes Gutenberg Universit{\"a}t, 55128 Mainz, Germany}
\affiliation{Graduate School Materials Science in Mainz, Staudingerweg 9, 55128 Mainz, Germany}
\author{Ar.~Abanov}
\affiliation{Department of Physics \& Astronomy, Texas A\&M University, College Station, Texas 77843-4242, USA}
\author{J. Sinova}
\affiliation{Institute of Physics, Johannes Gutenberg Universit{\"a}t, 55128 Mainz, Germany}
\affiliation{Institute of Physics ASCR, v.v.i, Cukrovarnicka 10, 162 00 Prag 6, Czech Republic}
\author{K. Everschor-Sitte}
\affiliation{Institute of Physics, Johannes Gutenberg Universit{\"a}t, 55128 Mainz, Germany}

\date{\today}

\begin{abstract}
The analysis of domain wall dynamics is often simplified to one dimensional physics. For domain walls in thin films, 
more realistic approaches require the description as two dimensional objects.  
This includes the study of vortices and curvatures along the domain walls as well as the influence of boundary effects.
Here we provide a theory in terms of soft modes that allows to analytically study the physics of extended domain walls and their stability.
 By considering irregular shaped skyrmions as closed domain walls, we analyze their plasticity and compare their dynamics with those of circular skyrmions. Our theory directly provides an analytical description of the excitation modes of magnetic skyrmions, previously only accessible by sophisticated micromagnetic numerical calculations and spectral analysis. These analytical expressions provide the scaling behaviour of the different physics on  parameters that experiments can test. 
 \end{abstract}


\maketitle
\section{Introduction}

The existence of distinguishable magnetic domains in ferromagnetic films~\cite{Bloch1932, Landau1935} is the basis for  magnetic memories devices.~\cite{Parkin2008} The boundary between two opposite magnetic domains corresponds to a domain wall (DW).~\cite{Slonczewski1973,Hayashi2006,Tatara2004,Yamanouchi2004} DWs belong to a class of textures called topological textures. Analytically in the low-energy limit, DWs are often treated as one dimensional objects that can be described by two soft modes.~\cite{Thiele1973, Slonczewski1973, Slonczewski1972,Schryer1974,Thiaville2005,Tatara2008,Beach2005,Atkinson2003} 
Experiments in thin films~\cite{Jiang2015,Emori2014,Yamanouchi2004,Benitez2015,Boulle2016} reveal however much richer configurations including curved DWs and vortex DWs,~\cite{LaBonte1969}  which in such a one dimensional picture are not captured. Therefore a two dimensional theory is needed.

In this paper, we consider DWs as strings described by a pair of fields of soft modes. We derive a theory that allows to treat the local dynamics of the DW and interactions with localized perturbations like impurities. Also the effects of external perturbations can be easily added to the formalism. 

Beyond DWs, magnetic skyrmions are promissing candidates for applications in spintronics.~\cite{Bogdanov1989,Muhlbauer2009,Heinze2011,Nagaosa2013} Skyrmions have a quantized topological charge and respond more efficiently to applied currents.\cite{Jonietz2010, Schulz2012} 
Due to their reduced size and particle-like behavior they have attracted significant interest in spintronics~\cite{Wolf2006} with applications as memory and logic devices.~\cite{Parkin2008, Fert2013, Zhang2015l, Allwood2005} 
Skyrmions are very much related to DWs in the sense at they can be viewed as closed DWs. Often, their theoretical description is based on a radially symmetric ansatz corresponding to a circular DW.~\cite{Feldtkeller1965,Ezawa2010,Kiselev2011,Rohart2013} 
The radial symmetry, however, is not a requirement for their existence. Other shapes have been oberved, for example, close to a defect\cite{Mueller2015, Romming2015} or by exciting their internal modes.~\cite{Romming2015,Seki2012,Woo2016,Lin2014, Makhfudz2012} The plasticity of the skyrmion shape has several experimental advantages, including the ability to avoid impurities\cite{Mueller2015, Romming2015} and increasing the response to external electrical currents.~\cite{Iwasaki2013cc,Litzius2016} Here we describe skyrmions as closed DWs without imposing shape-related symmetries.
Our formalism directly allows to study the dynamics of deformed magnetic skyrmions and their eigenmodes.

The paper is organized as follows. In Sec.~\ref{sec:DWstring} we present the effective description of a DW as a string. We describe the ansatz and obtain the effective energy for these topological textures. In Sec.~\ref{sec:DWDyn} we obtain an action that describes the dynamics in terms of the soft mode fields. In Sec.~\ref{sec:eqcldw} we consider closed DWs and analyze the dynamics of deformed skyrmions. In particular, we consider three examples, 
we calculate the dynamics of deformations (bumps) along skyrmions and examine the rotational and the breathing mode dynamics of deformed skyrmions.
In Sec.~\ref{sec:conclusion} we report our conclusions.

\section{Description of Domain wall strings}
\label{sec:DWstring}

We consider a ferromagnetic film with thickness $\tau$, much smaller than the DW width, and with interfacial Dzyaloshinskii-Moriya interaction (DMI).~\cite{Moriya1960,Dzyaloshinsky1958,Chen2013} The magnetization field is described as  $\vect{M} (\vect{r}) = M_s \vect m (\vect r) $, where $\vect m (\vect r)$ is a unit vector field and $M_{s}$ is the saturation magnetization. The micromagnetic Hamiltonian corresponds to
\begin{align}\label{eq:MicromagneticH}
\mathcal{H} =& \tau \int d^2 x \Bigg(\frac{J}{2}|\nabla\hat{\vect{m}}|^2 +K(1 - m_{z}^2)\notag\\
&+ D\left(m_{z} \nabla \cdot \vect{m}_{\perp} - \vect{m}_{\perp} \cdot \nabla  m_{z}\right)\Bigg),
\end{align}
where $\vect{m}_{\perp}$ is the vector of the in-plane components of the magnetization $m_{x}, m_{y}$. The term proportional to $J>0$ is the ferromagnetic exchange interaction.
It favors the alignment of the magnetization vectors. The term proportional to $K$ correspond to out-of-plane anisotropy interaction. It favors the magnetization pointing along the easy-axis, perpendicular to the plane of the film. The term proportional to $D$ is the interfacial DMI. 
For $D \geq 4\sqrt{JK}/\pi$ the ground state is given by a helix,~\cite{Rohart2013} which can be viewed as a periodic structure of DWs. In order to describe single DWs, we consider $D < \sqrt{JK}$. We neglect explicit terms related to demagnetization fields as they may be interpreted as a contribution to an effective anisotropy in thin films.~\cite{Draaisma1988} The Hamiltonian from Eq.~\eqref{eq:MicromagneticH} has a characteristic length, the DW width $\Delta = \sqrt{J/2K}$, and a characteristic energy density scale, $\mathcal{E} = \sqrt{JK}$.~\cite{Hubert1998} These two parameters define the scale for the dynamics of topological textures.

\begin{figure}
    \includegraphics[scale=0.31]{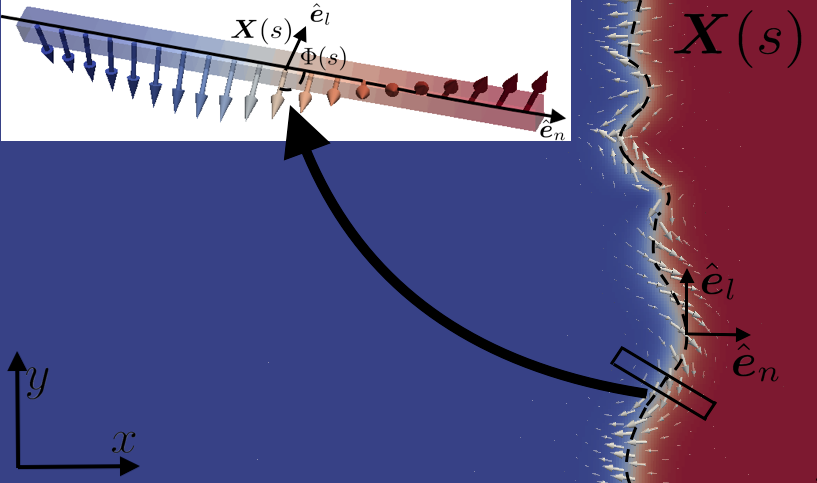}
    \caption{
        A sketch of an extended smooth DW given by the curve $\vect{X}(s)$ in a thin film. 
        Everywhere along the curve, the radius of curvature is much bigger than $\Delta$. The basis of the longitudinal, $\hat{\vect{e}}_{l}$, and normal, $\hat{\vect{e}}_{n}$, vectors are represented in the figure. On the top left corner, we show a typical cross section along the curve with the representations of $\vect{X}(s)$ and $\Phi(s)$
        \label{fig:eDW}
    }
\end{figure}

We consider a DW in a thin film as an extended object described by a curve $\vect{X}$, see Fig.~\ref{fig:eDW},
\begin{equation}\label{eq:DWcurve}
\vect{X}(s) = (x(s),y(s),0),
\end{equation}
where $s$ is a parameter along the curve.  A local basis along the curve is given by the unitary longitudinal vector $\hat{\vect{e}}_{l}(s)$ and the unitary normal vector $\hat{\vect{e}}_{n}(s)$
\begin{equation}
    \hat{\vect{e}}_{l} = \frac{\vect{X}'}{|\vect{X}'|}, \quad \hat{\vect{e}}_{n} = \hat{\vect{e}}_{l}\times \hat{\vect{e}}_{z},
\end{equation}
where for any function $f(s)$ we define $f' \equiv \partial_{s}f$ and $|\vect{X}'| = \sqrt{(\partial_{s}x)^2+(\partial_{s}y)^2}$.
Note that $\hat{\vect{e}}_{l}' = k\hat{\vect{e}}_{n}$, where the function $k\equiv k(s)$ is related to the local curvature $\kappa(s)$ via $k=|\vect{X}'| \kappa$, see App.~\ref{app:Ansatz}.
We assume that the radius of the local curvature $1/\kappa$ at any point along the DW is much bigger than the DW width, i.e.\ $|\vect{X}'|/k \gg \Delta$.

We consider an ansatz for an extended DW described by a curve $\vect{X}(s)$ with a cross section along $\hat{\vect{e}}_{n}$ corresponding to a rigid one-dimensional DW, see Fig.~\ref{fig:eDW} and App.~\ref{app:Ansatz}.
Thus, each cross section is described by a pair of soft modes $\vect{X}(s)$, the position of the cross section where the magnetization is in-plane, and $\Phi(s)$, the angle of the in-plane magnetization with respect to the vector $\hat{\vect{e}}_{n}(s)$. The effective description of the DW string is in terms of the fields $\vect{X}(s)$ and $\Phi(s)$.

In the case of finite thin films without periodic boundary conditions, DMI produces a twist of the magnetization at the boundaries.~\cite{Rohart2013, Meynell2014} This means that, for open DW curves in thin films, DMI twists the magnetic profile for cross sections close to the edge. We treat these edge twists as a perturbation to the state without edge twists and associate to them an effective potential. For DWs with periodic boundary conditions, there is no such twisting and, consequently, there is no potential associated to it.
The effective energy obtained from the micromagnetic Hamiltonian given by Eq.~\eqref{eq:MicromagneticH} with periodic boundary conditions is in terms of the effective coordinates given by

\begin{align}\label{eq:eDW}
\mathcal{H}_{\mathrm{eff}}= \tau \int ds \Big(&\mathit{c}_{\mathcal{E}}\mathcal{E}|\vect{X}'| + \frac{\mathit{c}_{\kappa}J}{|\vect{X}'|}\left(\Phi'-k\right)^2\notag\\
&+D \left(\pi|\vect{X}'|-\mathit{c}_{d}k \right)\cos\Phi\Big).
\end{align}
Here, $\mathit{c}_{\mathcal{E}},\mathit{c}_{\kappa}$ and $\mathit{c}_{d}$ are dimensionless constants that depend on the exact profile of the DW along the normal direction $\hat{\vect{e}}_{n}$, see Appendix~\ref{app:Ansatz}. This effective Hamiltonian is one of the main results of this paper. It is invariant under reparametrization, $s\rightarrow \gamma s$ where $\gamma \in \mathbb{R}$.
The first term in Eq.~\eqref{eq:eDW} is proportional to the length of the DW in analogy to the energy of a rubber band.
The solution that minimizes this term is a straight line.
The second term describes the fact that bending the DW leads to a changes in $\Phi'$ and vice versa. 
As the azimuthal angle can be manipulated experimentally for example by local external magnetic fields or spin waves,~\cite{Garcia-Sanchez2015,Wagner2016,Klaui2005} this provides a mechanism to introduce curvature in DWs.
Note that without DMI the energy is invariant under global rotations of the azimuthal angle of the DW. 
As DMI breaks inversion symmetry, it directly couples the azimuthal angle $\Phi$ with the curvature and the total length of the DW curve making the physics of the extended DW more complex.

To calculate the effective potential due to boundary conditions for open DWs, we assume the following two conditions: i) the longitudinal vector $\hat{\vect{e}}_{l}$ at the boundary is perpendicular to the edge; and ii) 
the length scale given by the DMI induced boundary twists is not much bigger than the DW width $\Delta$. The first condition implies that the magnetization configuration at the edge corresponds to a DW with the same width as in the bulk. 
Due to our assumption that the local curvature of the extended DW is larger than its width, the second condition grants that the DW is straight in the region close to the boundary. As a result, the DMI induced boundary condition leads to a rigid twist of the magnetization profile around $\hat{\vect{e}}_{n}$ for cross sections close to the edge.
As we substitute this modified ansatz for the edges into the micromagnetic Hamiltonian from Eq.~\eqref{eq:MicromagneticH}, we obtain that the energy potential due to edge effects does, as expected, not depend on the position $\vect{X}$. The main contributions are functions of $\Phi$ and $\Phi'$ defined only on a small region around the edges.
Within the above formalism, the Zeeman interaction due to an external magnetic field can also be incorporated. In this case the boundary condition will depend also on the relative positions of the string ends.~\cite{Boulle2013,Muratov2017}
In the following we will restrict our analysis to periodic boundary conditions. 

\begin{figure}
   \includegraphics[scale=0.32]{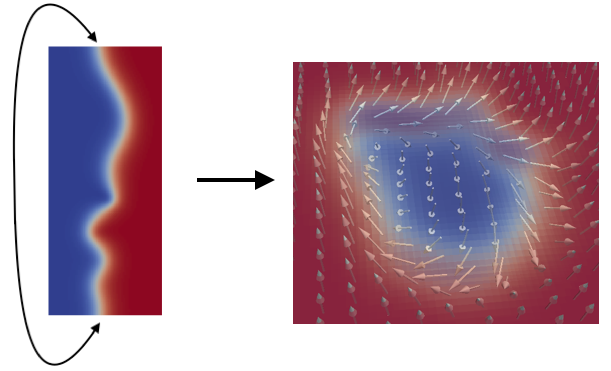}
    \caption{
        A sketch of how to obtain a skyrmion from extended DW with periodic boundary condition.
        \label{fig:ASky}
    }
\end{figure}

For 2d magnetic textures it is possible to define a topological charge of the form~\cite{Nagaosa:2013cc}
\begin{equation}\label{eq:topcharge}
    Q = \frac{1}{4\pi}\int d^2x\, \hat{\vect{m}}\cdot\left(\partial_{x}\hat{\vect{m}}\times\partial_{y}\hat{\vect{m}}\right). 
\end{equation}
The conservation of topological charge in the continuous approximation, is guaranteed  by the boundary conditions. In a magnetic lattice the conservation of topological charge holds only for textures that are bigger than the discretization scale. 
By substituting the ansatz of the DW string into Eq.~\eqref{eq:topcharge} we obtain,
\begin{equation}\label{eq:Tcharge}
Q = \frac{1}{2\pi} \int ds \left(k - \Phi' \right).
\end{equation}
For open DWs in thin films without periodic boundary conditions, the topological charge neither needs to be an integer number nor a conserved quantity. This means that it is possible to induce topological charge into these textures through boundary dynamics.~\cite{Muller2016,Du2015} In contrast, for DWs with periodic boundary conditions, this charge must be conserved in the continuum approximation as the integrals over $\Phi'$ and $k$ are quantized. 
Connecting the ends of a DW, see Fig.~\ref{fig:ASky}, leads to a skyrmion. For such closed DW strings without knots, the contribution from the integral over $k$ is $\pm 1$, where we use Gauss-Bonnet theorem. Thus, by interpreting skyrmions as closed DW strings we, of course, also obtain their quantized topological charge of $\pm1$ depending on their center magnetization. 
Eqs.~\eqref{eq:eDW} and \eqref{eq:Tcharge} reveal that a possible mechanism to produce skyrmions out of DW strings is to manipulate the azimuthal angle.~\cite{Milde2013}

In the following we will consider explicitly the dynamics of extended DWs in Sec.~\ref{sec:DWDyn} and the physics of closed DWs in Sec.~\ref{sec:eqcldw}.
For the case of open DWs the solution with minimum energy is given by a straight DW, where $|\vect X'|$ is a constant and $\Phi' = k = 0$;
while for a closed DW the state with minimum energy is given by a circular skyrmion with radius $R = \Delta\sqrt{\mathit{c}_{\mathcal{E}}/(\mathit{c}_{\kappa} - \pi D/\mathcal{E})}$ and a constant azimuthal angle $\Phi  = \pi$ along the curve.
Note that plugging in a circular skyrmion ansatz into Eq.~\eqref{eq:eDW} leads to the effective energy known from the literature for skyrmions.~\cite{Rohart2013}
 These two minimal solutions do not have local degrees of freedom for the soft mode fields. Thus, the long range dynamics of a straight DW is globally defined by the soft mode pair $\{X,\Phi\}$.~\cite{Rodrigues2017} In the case of skyrmions, also two globally defined soft modes are enough to capture their dynamics, $\{R,\Phi\}$, the radius of the skyrmion and the azimuthal angle of the in-plane magnetization along the radius.~\cite{Davi,Makhfudz2012}

\section{Effective dynamics of extended DWs}
\label{sec:DWDyn}

The model that best describes the magnetization dynamics in a ferromagnet is given by the Landau-Lifshitz-Gilbert (LLG) equation~\cite{Gilbert2004}
\begin{equation}\label{eq:LLG}
\dot{\hat{\vect{m}}} = \frac{\gamma}{M_{s}} \hat{\vect{m}} \times \frac{\delta \mathcal{H} [\hat{\vect{m}}]}{\delta \hat{\vect{m}}} + \alpha \hat{\vect{m}} \times \dot{\hat{\vect{m}}},
\end{equation}
where $\gamma$ is the gyromagnetic constant, $\mathcal{H}$ is the Hamiltonian of the system, $\alpha$ is the dimensionless Gilbert damping parameter, and we define $\dot{f}\equiv\partial_{t}f$ for any function $f$. The first term on the right-hand side corresponds to the precession of the magnetization due to an effective magnetic field and it conserves energy. The second term correspond to a damping term which promotes the alignment of the magnetization with the effective magnetic field.
The energy conserving part of the LLG Eq.~\eqref{eq:LLG} may be obtained from varying the action~\cite{Papanicolaou1991} 
\begin{align}\label{eq:Action}
    \mathcal{A} = \int dt\left(\frac{\tau M_{s}}{\gamma}  \int d^2x (1-\cos\theta)\dot{\phi} - \mathcal{H}\right),
\end{align}
where the first part is the spin Berry phase $S_{B}$ expressed in a standard spherical polar representation of the magnetization.~\cite{Braun1996}
 Plugging in Eq.~(\ref{eq:LLG}) the ansatz of a curved DW in terms of the fields of soft modes, we obtain
\begin{equation}\label{eq:BerryPhase}
    S_{B} = -\tau\mathit{c}_{\gamma} \int dt \int ds |\vect X'| (\dot{\vect{X}}\cdot\hat{\vect{e}}_{n})\left(\Phi - \arccos\left(\hat{\vect{e}}_{n} \cdot\hat{\vect{e}}_{x}\right)\right),
\end{equation}
where $c_{\gamma} = 2 M_{s} / \gamma$, see App. \ref{app:SB}.
From the spin Berry phase action, Eq.~\eqref{eq:BerryPhase}, and the effective Hamiltonian, Eq.~\eqref{eq:eDW}, it is possible to obtain the equations of motion for the soft mode fields $\vect{X}(s),\Phi(s)$ to study the dynamics of DWs in thin films, such as the propagation of extended DWs in different geometries, the influence of curvatures and the formation of cusps.
A full general description is however rather complicated as the  equations of motion are heavily influenced by the boundary conditions. 

In the case of $\Phi' = 0$, i.e.\ a constant azimuthal angle along the DW, the Berry phase from Eq.~\eqref{eq:BerryPhase} can be separated as
\begin{align}\label{eq:phiA}
S_{B} =& -\tau \mathit{c}_{\gamma} \int dt \, \Phi\dot{A}\notag\\
&+ \tau \mathit{c}_{\gamma} \int dt \int ds  |\vect X'| (\dot{\vect{X}}\cdot\hat{\vect{e}}_{n})\arccos\left(\hat{\vect{e}}_{n} \cdot\hat{\vect{e}}_{x}\right)\; ,
\end{align}
where by $A$ we define the area of the ferromagnetic domain with $m_z=-1$. The first term reveals that $A$ and $\Phi$ are conjugated soft modes,~\cite{Rodrigues2017} with a Poisson bracket given by $\{\Phi,A\} = 1/(\tau \mathit{c}_{\gamma})$. 
An important remark is that, since an external out-of-plane magnetic field couples directly to $A$, it produces, as expected, a precession of the angle $\Phi$.

\section{Effective dynamics of closed DWs}
\label{sec:eqcldw}
In this section we study the dynamics of closed DWs that can be parameterized within the polar coordinate representation
\begin{equation}\label{eq:curve}
\vect{X}(s,t) = r(s,t)(\cos(s),\sin(s),0),
\end{equation}
where $s = [0,2\pi)$, and $r(s)$ is a smooth function with $r(0,t)=r(2\pi,t)$.
Such an ansatz includes circular and distorted magnetic skyrmions as shown in the right panel of Fig.~\ref{fig:ASky}.
 For a curve given by Eq.~\eqref{eq:curve}, the effective spin Berry phase is 
\begin{equation}
S_{B} = - \tau \mathit{c}_{\gamma} \int dt \int_{0}^{2\pi} ds\,  \left(\Phi - \arctan\left(\frac{r'}{r}\right)\right)  r\dot{r}.
\end{equation}
The term depending on $\arctan\left(r'/r\right)$ is associated to the chiral properties of the skyrmion and is related to different speeds of propagation for waves in clockwise and counterclockwise motion.\cite{Makhfudz2012}

\subsection{Equations of motion for closed DWs}
The equations of motion for the local radial distance $r$ and the azimuthal angle $\Phi$ of the closed DW are
\begin{subequations}
\begin{align}
c_{\gamma} r\dot{r} =& \left(\frac{2J\mathit{c}_{\kappa}}{|\vect{X}'|}\left(\Phi' - k\right)\right)'+ D\sin\Phi \left(\pi |\vect{X}'| - \mathit{c}_{d}k\right),\label{eq:dr}\\
- \tau c_{\gamma} \dot{\Phi} =& \frac{1}{ r}\frac{\delta}{\delta r}\left(\mathcal{H}_{\mathrm{eff}} -  \tau c_{\gamma}r\dot{r}\arctan\left(\frac{r'}{r}\right)\right)\label{eq:phi},
\end{align}
\end{subequations}
where $|\vect{X}'| = \sqrt{r^2 + r'^2}$ and $k =  (\arctan(r'/r))'-1$. 
Here, we did not include damping dynamics. The addition of the Gilbert-damping term corresponds to adding a term proportional to the damping constant $\alpha$ that mixes the evolution of $\dot{r}$ and $\dot{\Phi}$, see App.\ref{app:Damping}.

Eq.~\eqref{eq:dr} has two interesting properties that we would like to point out: i) we find that 
the only contribution to the time evolution of the area enclosed by the DW is due to the DMI. In the case of  $\Phi' =0$ we obtain after integrating the equation of motion along the curve
\begin{equation}
\label{eq:dA}
    \dot{A} = \frac{\pi D}{\mathit{c}_{\gamma}}\sin\Phi\left(2 \mathit{c}_{d} +  \int_{0}^{2\pi} ds |\vect{X}'|\right).
\end{equation}
ii) In the absence of DMI, Eq.~\eqref{eq:dr} has the form of a continuity equation, i.e. $\dot{\rho} = \partial_s (j)$. 
Here,  analogous to the charge density, it is proportional to the area per unit length 
and $j$ corresponds to the density of topological charge per local length change. 
This shows that the existence of Bloch lines, which corresponds to a concentration of topological charge, can generate strong deformations of the DW.

Eq.~\eqref{eq:phi} is rather complicated as it contains higher order derivatives. Integrating the equation of motion along the entire curve gives
\begin{align}
&c_{\gamma}\int ds \left(\dot{\Phi}+ \frac{r^2\dot{r}'}{|\vect{X}'|^2}\right)\notag\\
&= \int ds \Bigg(\frac{2(\mathcal{E}\mathit{c}_{\mathcal{E}}+D\pi\cos\Phi)}{|\vect{X}'|} - \frac{J\mathit{c}_{\kappa}}{|\vect{X}'|^{3}}\left(\Phi' - k\right)^2\nonumber\\
&-\left(\frac{r'}{r|\vect{X}'|^2}\right)\left(
\frac{2J\mathit{c}_{\kappa}}{|\vect{X}'|}\left(\Phi' - k\right)+D\mathit{c}_{d}\cos\Phi\right)'\Bigg),
\label{eq:dphi}
\end{align}
where some of the terms with higher order derivatives vanished due to the periodic boundary conditions.

These equations provide a good intuition to the dynamics of extended DWs as well as  deformed skyrmions, and convey the general dynamics of a broad range of magnetic textures. In particular, this effective theory allows for the study of continuous deformations of skyrmions without requiring the knowledge of magnonic modes. 
In the following we will explicitly explore three examples, the propagation of deformations along a skyrmion, the eigenmodes of closed DWs and breathing dynamics.

\subsection{Wave propagation along the closed DW} 
Here we consider circular skyrmions with radius $R_{0}$ with a small and smooth deformation of size $d$, i.e.\  $r(s,t) = R_{0} + d(s,t)$ with $d\ll R_{0}$.
We assume that the angle $\Phi$ and the radius $R_{0}$ are fixed to the values minimizing the energy,\cite{Rohart2013} i.e.\ $\Phi = \pi$ and $R_{0}= \Delta\sqrt{\mathit{c}_{\kappa}/(\mathit{c}_{\mathcal{E}} - \mathit{c}_{\pi}(D/\mathcal{E})}$ where $\Delta$ is the DW width.
In this case we obtain from Eq.~\eqref{eq:dr} a drift equation 
\begin{align}
    \left(\frac{R_{0}^3\mathit{c}_{\gamma}}{2\mathit{c}_{\kappa}J}\dot{d} + d'\right) \approx 0.
\end{align}
This describes the propagation of the deformation $d$ along the DW. Since the curve is closed, this motion is periodic with frequency $\omega \propto J\gamma/M_{s}R_{0}^{3}$.

\subsection{Rotational eigenmodes of a closed DW} 
\begin{figure}
    \includegraphics[scale=0.25]{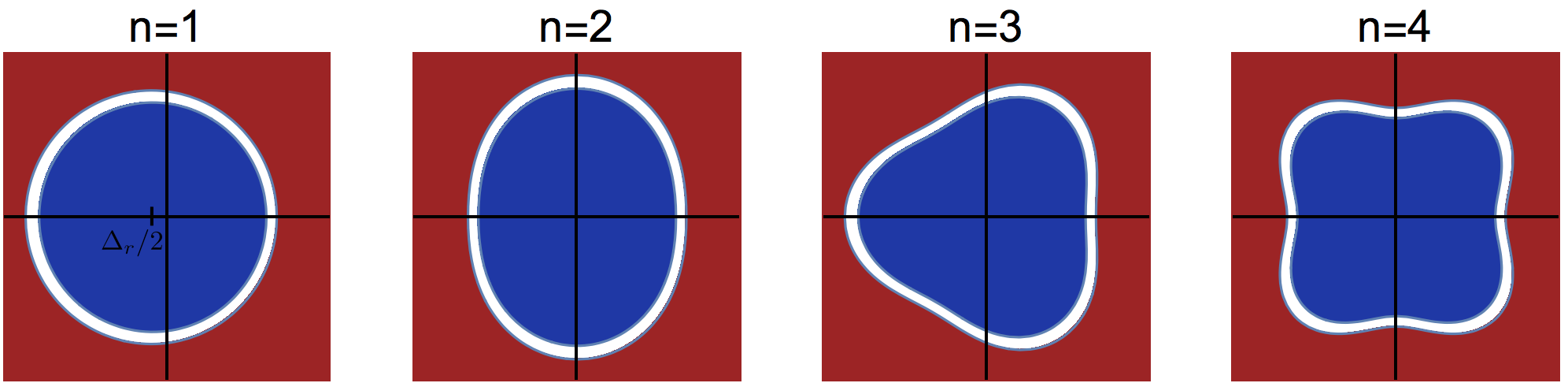}
    \caption{
    Sketch of solutions described by Eq.~\eqref{eq:ShapeSky} with $\Delta_{r}/r_{\mathrm{min}}= 0.3$ and $n=1,2,3,4$. 
    These solutions correspond to the excitations of circular skyrmions described in the literature.~\cite{Lin2014, Makhfudz2012,Schutte2014,Sheka2001, Kravchuk2017a}
    Their evolution in time is a rigid rotation with a frequency depending on their size. Notice that the mode $n=1$, corresponds to what is known as (counter-) clockwise modes for skyrmions in the literature.  
        \label{fig:HMod}
    }
\end{figure}

In this example we consider solutions of Eq.~\eqref{eq:dr} that have a rigid shape and obey the following conditions:
i) the curve has a maximum $r_\mathrm{max}$ and a minimum $r_\mathrm{min}$ radius, which are of the order of the equilibrium radius $R_0$ of a circular skyrmion; ii) the solution depends on the combination $s - \omega t$, where $s$ is the parameter along the curve, $t$ is time, and $\omega$ is a constant frequency; iii) the solution is periodic; and iv) $\Phi'=0$. With these conditions, we find that the curve describing the DW string, is given by
\begin{equation}\label{eq:ShapeSky}
    r(s) \approx r_{\mathrm{min}} + \Delta_{r}\sin^2\left(\frac{n s}{2}\right),
\end{equation}
where $\Delta_{r} = r_{\mathrm{max}}-r_{\mathrm{min}}$, and $n$ is an integer. 
The first four solutions of the above equation are shown in Fig.~\ref{fig:HMod}.  The solutions in Eq.~\eqref{eq:ShapeSky}, which we obtained within our effective theory, correspond to the skyrmion excitation modes reported in the literature.~\cite{Lin2014,Makhfudz2012,Schutte2014,Sheka2001,Kravchuk2017a}
The strength of our formalism is that we can analytically obtain the frequencies.
The frequency $\omega_n$ for the n-th mode is given by
\begin{align}
    \omega_{n} \approx - \frac{2J\mathit{c}_{\kappa}}{\mathit{c}_{\gamma}r_{\mathrm{min}}^3}(n - n^3).
\end{align}
This result is in agreement with the result obtained by numerical methods by \textit{ Kravchuk et al.\ } for the clockwise modes.~\cite{Kravchuk2017a}
An important remark is that, even though this frequency does not depend explicitly on the DMI, this interaction is necessary to keep $\Phi$ constant along the DW string. 

To obtain the full spectrum, we need also to consider the cases in which $\dot{\Phi} \neq 0$. For small perturbations of $r$ and $\Phi$, around the minimum radius $r_{\mathrm{min}}$ and the equilibrium angle $\pi$, Eqs.~\eqref{eq:phi} and \eqref{eq:dr} become
\begin{align}
c_{\gamma}\dot{r} + D\pi \Phi &\approx \mathcal{O}\left(\frac{1}{r_{min}^3}\right)\notag\\
c_{\gamma}\left(r_{\mathrm{min}}\dot{\Phi} -2\partial_{\psi}\dot{r}\right)&\approx \mathcal{O}\left(\frac{1}{r_{min}^2}\right).
\end{align}
These equations of motion correspond to oscillations with frequency
\begin{equation}
\omega_{n} \approx -\frac{2\pi D n}{c_{\gamma} r_{\mathrm{min}}},
\end{equation}
which corresponds to the counter-clockwise modes reported by \textit{ Kravchuk et al}.~\cite{Kravchuk2017a} To understand the difference for the clockwise and counterclockwise modes, we notice that for the clockwise modes, $n < 0$, oscillations for $\Phi$ and $r$ travel in the same direction while for the counterclockwise modes, $n>0$, they travel in opposite directions. Since in this formalism, the angle $\Phi$ is defined with respect to the normal direction along the DW string, an oscillation for $r$ already assumes a rotation of the local magnetization. Therefore, if both oscillations travel in the same direction, the oscillation in $r$ can suppress the oscillation for $\Phi$, remaining the frequency $\sim 1/r_{\textrm{min}}^3$, while if they travel in opposite directions, the oscillation in $\Phi$ should be increased, producing the motion with frequency $\sim1/r_{\textrm{min}}$.

\subsection{Breathing modes of a closed DW}

Here we consider the smooth breathing dynamics of a closed DW with $\Phi'=0$ along the curve. This means that its shape is fixed and the dynamics is given by a scaling factor $\epsilon(t)$, i.e.\ $r(t,s) = (1 + \epsilon(t))r_{0}(s)$, where $r_{0}(s)$ is the equilibrium fixed shape. From Eqs.~\eqref{eq:dA} and \eqref{eq:dphi}, we obtain the following equations in terms of the scalling factor $\epsilon(t)$,
\begin{subequations}
\begin{align}
    \dot{\epsilon} &= \frac{\pi D}{2A_{0} \mathit{c}_{\gamma}} \sin\Phi\left(\int_{0}^{2\pi} ds |\vect{X}_{0}'| + 2 \mathit{c}_{d}(1-\epsilon)\right),\\ 
    2\pi c_{\gamma}\dot{\Phi} &= \left(\int ds \frac{2(\mathcal{E}\mathit{c}_{\mathcal{E}} + D\cos\Phi)}{|\vect{X}_{0}'|} - \int ds \frac{J\mathit{c}_{\kappa}k_{0}^2}{|\vect{X}_{0}'|^3}\right)\notag\\
    & - \epsilon \left(\int ds \frac{2(\mathcal{E}\mathit{c}_{\mathcal{E}} + D\pi\cos\Phi)}{|\vect{X}_{0}'|} - 3\int ds \frac{J\mathit{c}_{\kappa}k_{0}^2}{|\vect{X}_{0}'|^3}\right),
\end{align}
\end{subequations}
where $A_{0}$, $|\vect{X}_{0}'|$ and $k_{0}$ are only functions of the initial curve of the DW given by $r_{0}(s)$. Note that the second term in Eq.~\eqref{eq:dphi} is zero for the higher modes, due to symmetries. Also, in the case of smooth breathing dynamics the last term in Eq.~\eqref{eq:dphi} is small and therefore is neglected.
From the first equation and first term in the second equation we obtain the condition for the equilibrium configuration. They fix $r_{0}(s)$ and $\Phi = \pi$.

We define $r_{0}(s) = \tilde{R}_{0}\mathrm{r}(s)$, where $\mathrm{r}(s)$ is dimensionless and $\mathrm{r}(s) = 1$ at the average radius $\tilde{R}_0$. The equilibrium condition fixes $\tilde{R}_{0}$ for any deformed skyrmion to\begin{equation}
\tilde{R}_{0} = \Delta\sqrt{\frac{\tilde{\mathit{c}}_{\kappa}}{\tilde{\mathit{c}}_{\mathcal{E}} - \tilde{\mathit{c}}_{\pi}(D/\mathcal{E})}},
\end{equation}
where $\Delta$ is the DW width and we have rescaled the constants $\tilde{\mathit{c}}_{\pi} = \pi\int ds \left(\mathrm{r}_{0}^2 + \mathrm{r}_{0}'^2\right)^{-1/2}$, $\tilde{\mathit{c}}_{\mathcal{E}} = \mathit{c}_{\mathcal{E}}\int ds \left(\mathrm{r}_{0}^2 + \mathrm{r}_{0}'^2\right)^{-1/2}$ and 
$\tilde{\mathit{c}}_{\kappa}=\mathit{c}_{\kappa}\int ds k_{0}^2\left(\mathrm{r}_{0}^2 + \mathrm{r}_{0}'^2\right)^{-3/2}$.
This equilibrium radius is analogous to the case of a circular skyrmion with rescaled parameters.~\cite{Rohart2013}

If we consider small perturbations of $\Phi$ around the equilibrium configuration of the skyrmion, $\Phi = \pi + \varphi$, we obtain the following equations of motion
\begin{subequations}
\begin{align}
    \dot{\epsilon} &= -\varphi\frac{\pi D}{2A_{0} \mathit{c}_{\gamma}}\left(\int_{0}^{2\pi} ds |\vect{X}_{0}'| + 2 \mathit{c}_{d}\right),\\ 
    \dot{\varphi} &=   \epsilon \int ds \frac{J\mathit{c}_{\kappa}k_{0}^2}{\pi c_{\gamma}|\vect{X}_{0}'|^3},
\end{align}
\end{subequations}
describing the breathing dynamics of deformed skyrmions with frequency 
\begin{equation}
\omega^2 = \frac{DJ \tilde{\mathit{c}}_{\kappa}}{2A_{0} \mathit{c}_{\gamma}^2}\left(\int_{0}^{2\pi} ds |\vect{X}_{0}'| + 2 \mathit{c}_{d}\right).
\end{equation}
In case of a circular skyrmion, this breathing mode corresponds to the zero order excitation mode.~\cite{Makhfudz2012,Sheka2001,Kravchuk2017a} Taking into account the dependence on the average radius $R_{0}$ of $\tilde{c}_{\kappa}, A_{0}$ and the total length of the domain wall, we notice that the frequency dependence on the radius for this mode is $\omega \sim 1/R_{0}^2$ for the first term.~\cite{Sheka2001,Kravchuk2017a} The second term, proportional to $c_{d}$, introduces a modification to the frequency with dependence $\sim 1/R_{0}^{5/2}$.

\section{Discussion and Conclusion}
\label{sec:conclusion}

Motivated by the experimentally observed richer configurations of topological magnetic textures in thin films, in this paper we extended the soft mode formalism of DWs beyond the effective one dimensional theory. We provided an ansatz for the effective description of a DW in terms of soft mode fields.  From the micromagnetic Hamiltonian for ferromagnetic thin films we obtained an effective energy for these topological textures revealing analogies of the DW string to elastic rubber bands. An important remark is that our results are invariant under reparametrization. We have shown that circular skyrmions are (meta-)stable configurations of the Hamiltonian and analyzed their stability. We calculated the effective spin Berry phase from which, associated with the effective energy, and obtain the equations of motion describing the dynamics of these structures. Finally we have applied our formalism to three examples of skyrmion dynamics. This allowed us to calculate the excitation modes of magnetic skyrmions analytically in agreement with previous works which only obtained them by sophisticated micromagnetic numerical calculations and spectral analysis. Our analytical expressions provide the scaling behaviour of the frequencies on different measurable parameters. 

We would like to emphasize that our developed formalism takes into account several aspects of topological textures dynamics, such as curvatures and variations of the in-plane magnetization. 
We have derived the topological charge for the extended DWs. Our theory allows the study of more complex dynamics like the propagation of extended DWs in different geometries with inhomogeneous perturbations, the local interaction of skyrmions and DWs, as well as the the annihilation and creation of skyrmions mediated by DWs. 
Furthermore, it is possible to analyze the stability of Bloch lines~\cite{Slonczewski1975,Jantz1981,DaCol2014} in DWs and the creation of skyrmion lattices by excitation of worm domains.~\cite{Jiang2015,Koshibae2014}

For a more general case, however, a full treatment of the boundary contributions is required.\cite{Rohart2013} These boundary conditions can be treated effectively\cite{Muratov2017} and lead to additional terms in Eq.~\eqref{eq:eDW}. Micromagnetic simulations show that the edges may be sources of both curvature and Bloch lines,\cite{Muller2016a} and with the presented formalism we can analyze their propagation along the DW.

We conclude by mentioning that our formalism is extendible with regard to different types of interactions or more generalized types of DMI,\cite{Hoffmann2017, Hals2017} which will cover the dynamics of more complex topological textures like antiskyrmions.

%%%%%%%%%%%%%%%%%%%%%%%%%%%%%%%%%%%

\section{Acknowledgments}
During the preparation of the manuscript we have learned that Oleg
Tchernyshyov and Shu Zhang have independently developed a similar
formalism.~\cite{Oleg} We thank the authors for the discussions. We also thank Olena Gomonay, Benjamin McKeever and Volodymyr Kravchuk for helpful discussions.
D.R.R.\ is a recipient of a DFG fellowship/DFG-funded
position through the Excellence Initiative by the Graduate School Materials Science in Mainz (GSC 266). J. S. acknowledges funding from the Alexander von Humboldt Foundation, the Transregional Collaborative Research Center (SFB/TRR) 173 SPIN+X, and Grant Agency of the Czech Republic grant no. 14-37427G.
 K.E.-S.\ acknowledges funding from the German Research Foundation (DFG) under the Project No. EV 196/2-1.

%%%%%%%%%%%%%%%%%%%%%%%%%%%%%%%%%%%%%%%%%%%%%%%
\appendix
\section{Derivation of the effective energy for the extended DW}\label{app:Ansatz}
In this paper we describe an extended DW in terms of a curve
$\vect{X}(s) = (x(s),y(s),0)$ and a local unitary basis along the DW given by the longitudinal, $\hat{\vect{e}}_{l}$, and the normal, $\hat{\vect{e}}_{n}$, vectors to the curve. We make the ansatz that the DW can be described along the normal direction by a rigid one-dimensional DW. 
We can transform the Cartesian basis $\hat{\vect{e}}_x,\hat{\vect{e}}_y$ via a local rotation into the basis $\hat{\vect{e}}_{n},\hat{\vect{e}}_{l}$ along the curve
\begin{align}
    x,y \, \rightarrow \, n,l,
\end{align}
where $n, l$ are the coordinates along the vectors $\hat{\vect{e}}_{n},\hat{\vect{e}}_{l}$, respectively, 
\begin{equation}\label{eq:nl}
n(s) = (\vect{x}-\vect{X}(s)) \cdot \hat{\vect{e}}_{n}(s), \quad l(s) = (\vect{x}-\vect{X}(s))\cdot \hat{\vect{e}}_{l}(s).
\end{equation}
Here $\vect{x}$ denotes any position in space
and the parameter $l$ takes values in between zero and the length of the extended DW.  
The transformation between the coordinate $l$ and any string parameter $s$ is given by
\begin{equation}\label{eq:UnitLength}
    dl = ds |\vect{X}'|.
\end{equation}
The curvature $\kappa$ of the DW is defined by~\cite{DoCarmo1976}
\begin{equation}
\partial_l \hat{\vect e}_l \equiv \kappa \hat{\vect e}_n = \frac{k}{|\vect{X}'|} \hat{\vect e}_n,
\end{equation}
where in the last equality we have used Eq.~\eqref{eq:UnitLength}.
According to our ansatz, we separate the micromagnetic Hamiltonian from Eq.~\eqref{eq:MicromagneticH} into three terms, 
\begin{align}
\mathcal{H} = \mathcal{H}_{n} + \mathcal{H}_{l} + \mathcal{H}_{D}
\end{align}
where
\begin{subequations}
\label{eq:H}
\begin{align}
\mathcal{H}_{n} &= \tau\int dl \int dn \left(\frac{J}{2}\left(\partial_{n}\hat{\vect{m}}\right)^2 +K(1 - m_{z}^2)\right), \\
\mathcal{H}_{l} &= \tau\int dl \int dn \left(\frac{J}{2}\left(\partial_{l}\hat{\vect{m}}\right)^2\right), \\
\mathcal{H}_{D} &= \tau\int dl \int dn D\Big(m_{z}(\hat{\vect e}_{n}\partial_{n} + \hat{\vect e}_{l}\partial_{l}).(m_{n}\hat{\vect e}_{n} + m_{l}\hat{\vect e}_{l})\notag\\
&- (m_{n}\partial_{n} +m_{l}\partial_{l})m_{z}\Big).
\end{align}
\end{subequations}
are the normal, longitudinal and DMI contribution to the Hamiltonian respectively. 
We now show that our ansatz, where each cross section corresponds to a rigid one-dimensional DW, minimizes $\mathcal{H}_{n}$ with the given DW boundary conditions. This can be directly seen by rewriting the exchange part in a spherical representation of the magnetization

\begin{align}
\mathcal{H}_{n} &= \frac{\tau\sqrt{JK}}{\sqrt{2}} \int dl \int d\tilde{n} \Big(\left(\partial_{\tilde{n}}\theta\right)^2 + \sin^2\theta(\partial_{\tilde{n}}\Phi)^2\notag\\
& +(1 - m_{z}^2)\Big),
\end{align}
where $\tilde{n} = n/\Delta$ is dimensionless. The state with minimum energy and DW boundary conditions, $\theta(n_{\mathrm{left}}) = \pi$ and $\theta(n_{\mathrm{right}}) =  0$, corresponds to a solution with $\partial_{\tilde{n}}\Phi = 0$ and a specific $\theta(\tilde{n})$ dependence. This solution is invariant under a rigid rotation of $\Phi$.
Furthermore, in the linear approximation $\theta$ does not depend explicitly on $l$.
The magnetization is then given by
\begin{align}\label{eq:mpt}
\hat{\vect{m}} &=\cos\Phi(l)\sin\theta(n) \hat{\vect{e}}_{n}(l) + \sin\Phi(l)\sin\theta(n) \hat{\vect{e}}_{l}(l)\notag\\
&+ \cos\theta(n) \hat{\vect{e}}_{z},
\end{align}
where $\theta(n(s))$ corresponds to a rigid DW centered at position $\vect{X}(s)$. Plugging the magnetization configuration of Eq.~\eqref{eq:mpt} 
into Eq.~\eqref{eq:H} and perfoming the integration over $n$ gives
\begin{subequations}
\label{eq:Hprime}
\begin{align}
\mathcal{H}_{n} &= \tau \mathit{c}_{\mathcal{E}} \sqrt{JK} \int dl \\
\mathcal{H}_{l} &= \tau J\mathit{c}_{\kappa}\int dl  \frac{1}{|\vect{X}'|^2}\left(\Phi'-k\right)^2, \\
\mathcal{H}_{D} &= \tau D \int dl \left(\pi -\mathit{c}_{d}\frac{k}{|\vect{X}'|} \right)\cos\Phi,
\end{align}
\end{subequations}
where $\mathit{c}_{\mathcal{E}}, \mathit{c}_{\kappa}, \mathit{c}_{d}$ are dimensionless constants depending on the precise shape of the DW.
Note that the DMI part of the Hamiltonian $\mathcal{H}_D$ has the following boundary dependent azimuthal angle contribution $\tau D 
\mathit{c}_d \int (dl/ |\vect X'|) \Phi' \cos\Phi$. Upon integration along the length of the DW curve this term vanishes for periodic boundary conditions. Moreover, for systems in which the domain wall profile is point symmetric to respect to the center of the domain wall, we have that $\mathit{c}_{d} = 0$. This point symmetry, however, is not always satisfied and can be broken, for example, by an out-of-plane magnetic field.

\section{Spin Berry phase in the extended DW ansatz}
\label{app:SB}

The spin Berry phase in a standard spherical representation of the magnetization
in thin film is~\cite{Braun1996} 
\begin{align}\label{eq:Berryphase}
    S_{B} = \frac{\tau M_{s}}{\gamma}\int dt \int d^2x (1-\cos\theta)\dot{\phi}.
\end{align}
 Performing a partial integration we obtain 
 \begin{equation}\label{eq:Berryphase1}
    S_{B} = -\frac{\tau M_{s}}{\gamma}\int dt \int d^2x \, \phi \, \sin\theta\, \dot\theta .
\end{equation}
This representation has the advantage that the non-vanishing contributions come only from regions within the DW, as $\sin\theta = 0$ for the ferromagnetic domains.
 For the time variation of $\theta$, we note that $\theta\equiv \theta(n)$. Any variation of $\theta$ correspond to a translation along the local $\hat{\vect{e}}_{n}$ direction. Using Eq.~\eqref{eq:nl}, we get
\begin{equation}
\dot{\theta} = - (\dot{\vect{X}} \cdot\hat{\vect{e}}_{n}) \partial_{n}\theta.
\end{equation}
Plugging this into Eq.~\eqref{eq:Berryphase1} and changing to the $n,l$ system of coordinates gives
\begin{equation}
    S_{B} = -\frac{\tau M_{s}}{\gamma}\int dt \int dn\, dl \left(\phi\, (\dot{\vect{X}} \cdot \hat{\vect{e}}_{n}) \partial_{n}\cos\theta\right).
\end{equation}
Since $\theta(n)$ is fixed by the DW solution, we may integrate the $n$ dependence and we obtain
\begin{equation}\label{eq:Berryphase2}
    S_{B} = - \tau c_{\gamma}\int dt \int dl \,\phi \,(\dot{\vect{X}} \cdot\hat{\vect{e}}_{n}) ,
\end{equation}
where $c_{\gamma} = 2 M_{s} / \gamma$ and with the reparametrization from $l$ to $s$ we obtain Eq.~\eqref{eq:BerryPhase} of the main text.
The magnetization azimuthal angle $\phi$ in terms of the DW coordinates is
\begin{equation}\label{eq:phi}
\phi = \Phi +\psi - \arccos\left(\hat{\vect{e}}_{n} \cdot\hat{\vect{e}}_{x}\right).
\end{equation}
Notice that the integral of the term proportional to $\psi$ in the Berry phase corresponds to a total time derivative and does not contribute to the dynamic equations. With Eqs.~\eqref{eq:Berryphase2} and \eqref{eq:phi} we obtain, therefore the effective Berry phase in Eq.~\eqref{eq:BerryPhase} in the main text. Also, as a remark, for a circular skyrmion we have that $\arccos\left(\hat{\vect{e}}_{n}\cdot\hat{\vect{e}}_{x}\right) = \psi$. 

A simpler case is when $\partial_{l}\Phi = 0$, meaning that $\Phi$ is globally defined along the DW curve. In this case, the first term for the effective Berry phase \eqref{eq:BerryPhase} becomes
\begin{equation}\label{eq:Phiconst}
S_{B1} = - \tau c_{\gamma}\int dt\,  \Phi \int dl (\dot{\vect{X}} \cdot\hat{\vect{e}}_{n}).
\end{equation}
We can further simplify this equation by the following considerations:
The infinitesimal area spanned by $dn$ and $dl$ is given by $dA = \left(dn \hat{\vect{e}}_n\times dl \hat{\vect{e}}_{l} \right)\cdot \hat{\vect e}_z.$ 
A time variation of this area corresponds to $dn \rightarrow dn +\delta_n$ and $dl \rightarrow dl +\delta_l$, such that 
\begin{equation}
\delta dA = \left(\delta_n\hat{\vect{e}}_n\times dl \, \hat{\vect{e}}_{l} \right)\cdot \hat{\vect e}_z
+\left(dn \hat{\vect{e}}_n\times \delta_l \, \hat{\vect{e}}_{l} \right)\cdot \hat{\vect e}_z.
\end{equation}

We want to study the area change upon changing the DW string $\vect X(t)$ with a fixed parameterization, i.e. $\dot{dA}$. 
\begin{equation}
\dot{d A} = \left((\dot{\vect{X}} \cdot \hat{\vect{e}}_{n})
\hat{\vect{e}}_n\times dl \, \hat{\vect{e}}_{l} \right)\cdot \hat{\vect e}_z=
dl (\dot{\vect{X}} \cdot \hat{\vect{e}}_{n})
\end{equation}
Note, that the second term vanishes as $\delta_l=0$ for a fixed parameterization and in the first term we have calculated the variation of the DW string along the normal direction.

Therefore, the second integral in Eq.~\eqref{eq:Phiconst} becomes
\begin{equation}
\int dl (\dot{\vect{X}} \cdot\hat{\vect{e}}_{n})
= \int \dot{(d A)}
= \dot{A},
\end{equation}
and we obtain the first term in Eq.~\eqref{eq:phiA} of the main text.

\section{Damping terms and Rayleigh functional}
\label{app:Damping}

The damping term in the Landau-Lifshitz-Gilbert equation, Eq.~\eqref{eq:LLG} of the main text,
can be derived from a Rayleigh functional.~\cite{Gilbert2004} This means that we can identify
\begin{equation}
\alpha \hat{\vect{m}} \times \dot{\hat{\vect{m}}} \leftrightarrow \frac{\delta R[\dot{\hat{\vect{m}}}]}{\delta \dot{\hat{\vect{m}}}},
\end{equation}
where the Rayleigh functional $R[\hat{\vect{m}}]$ is given by
\begin{equation}
\mathcal{R} [\dot{\hat{\vect{m}}}] = \frac{\alpha \tau M_{s}}{2\gamma}\int d^2x (\dot{\hat{\vect{m}}})^2.
\end{equation}
If we plug in the ansatz of a DW curve, we obtain
\begin{align}
\mathcal{R} [\dot{\hat{\vect{m}}}] &= \frac{\alpha \tau M_{s}}{2\gamma}\int d^2x \left(\sin^2\theta\, (\dot{\Phi})^2 + (\dot{\theta})^2
\right)\notag\\
&= \frac{\alpha \tau M_{s}}{2\gamma}\int dn\, dl \left(\sin^2\theta\, (\dot{\Phi})^2 + (\partial_{n}\theta)^2(\dot{\vect{X}}\cdot \hat{\vect{e}}_{n})^2
\right)\notag\\
&= \frac{\alpha \tau M_{s}}{2\gamma}\int dl \left(\mathit{c}_{\Phi}(\dot{\Phi})^2 + c_{X}\left(\dot{\vect{X}}\cdot \hat{\vect{e}}_{n})^2
\right)\right).
\end{align}
As the only non vanishing contribution comes from within the DW curve, we can change the $x,y$ basis to the local $n,l$ basis and then integrate 
the expression along the normal direction. The constants $\mathit{c}_{\Phi}, \mathit{c}_{X}$ are positive and dimensionless. They depend only on the specific shape of the profile.

The effect of damping occurs as additional terms $\gamma_{r}, \gamma_{\Phi}$ in the equations of motion of the closed DW, Eq.~\eqref{eq:dr} of the main text
\begin{subequations}
\begin{align}
c_{\gamma} r\dot{r} =& \left(\frac{2J\mathit{c}_{\kappa}}{|\vect{X}'|}\left(\Phi' - k\right)\right)'\\&+ D\sin\Phi \left(\pi |\vect{X}'| - \mathit{c}_{d}k\right) +  \gamma_{r}/\tau,\\
- \tau c_{\gamma} \dot{\Phi} =& \frac{1}{ r}\frac{1}{ r}\frac{\delta}{\delta r}\left(\mathcal{H}_{\mathrm{eff}} -  \tau c_{\gamma}r\dot{r}\arctan\left(\frac{r'}{r}\right)\right) +\gamma_{\Phi}.
\end{align}
\end{subequations}
Here $\gamma_r$ and $\gamma_{\Phi}$ for the curve of Eq.~\eqref{eq:curve} are given by
\begin{subequations}
\begin{align}
\gamma_{r} &= \frac{\alpha \tau M_{s}}{2\gamma}c_{\Phi}\dot{\Phi},\\
\gamma_{\Phi} &= -\frac{1}{r}\frac{\alpha \tau M_{s}}{2\gamma}c_{X}\dot{r}.
\end{align}
\end{subequations}
This reveals that the effect of damping is to couple the dynamics of $\dot{r}$ and $\dot{\Phi}$. To obtain the above damping terms, we followed the Thiele approach,~\cite{Thiele1973,Davi} where
\begin{subequations}
\begin{align}
\gamma_{r} &= \delta_{\dot{\Phi}}\mathcal{R} [\dot{\vect{m}}]/\int d^2x \left( \hat{\vect{m}}\cdot \partial_{\Phi}\hat{\vect{m}} \times \partial_{r}\hat{\vect{m}}\right),\\
\gamma_{\Phi} &= \delta_{\dot{r}}\mathcal{R} [\dot{\vect{m}}]/\int d^2x \left( \hat{\vect{m}}\cdot \partial_{r}\hat{\vect{m}} \times \partial_{\Phi}\hat{\vect{m}}\right).
\end{align}
\end{subequations}

\end{document}